\documentstyle[preprint,aps]{revtex}
\tighten
\begin{document}
\draft
\preprint{Preprint Numbers: \parbox[t]{48mm}
        {ANL-PHY-7663-TH-93
\\UNITUE-THEP-13/1993}}
\title{Pion Loop Contribution to the \\Electromagnetic Pion Charge Radius}
\author{Reinhard Alkofer\footnotemark[1], Axel Bender\footnotemark[1] 
and Craig D. Roberts\footnotemark[2]}
\address{\footnotemark[1]Institut f\"{u}r Theoretische Physik,
Universit\"{a}t T\"{u}bingen, \\
Auf der Morgenstelle 14, D-72076 T\"{u}bingen, FRG\\ 
\footnotemark[2]Physics Division, Argonne National Laboratory,\\
Argonne, Illinois 60439-4843, USA}
%
\maketitle
\begin{abstract}
A phenomenological Dyson-Schwinger equation approach to QCD, formalised in
terms of a QCD based model field theory, is used to calculate the
electromagnetic charge radius of the pion.  The contributions from the quark
core and pion loop, as defined in this approach, are identified and compared.
It is shown explicitly that the divergence of the charge radius in the chiral
limit is due to the pion loop and that, at the physical value of the pion
mass, this loop contributes less than 15\% to $\langle r^2_\pi\rangle$; i.e.
the quark core is the dominant determining characteristic for the pion.  This
suggests that quark based models which fail to reproduce the $\ln\,m_\pi$
divergence of $\langle r^2_\pi\rangle$ may nevertheless incorporate the
dominant characteristic of the pion: its quark core.
\end{abstract}
%
\section{Introduction}
Chiral symmetry and its dynamical breakdown is crucial to the nature of the
strong interaction spectrum and the study of this in Quantum Chromodynamics
(QCD), and models thereof, is of great current interest.  The most striking
signals of Dynamical Chiral Symmetry Breaking (DCSB) are found in the
kinematic and dynamical properties of the pion and one tool that has been
used extensively to study these properties is Chiral Perturbation Theory
(ChPT)\cite{EV81,GL83,GL85a,GL85}.  Chiral Perturbation Theory can be
understood as the study of the necessary consequences of the chiral Ward
Identities via the construction of an effective action, using field variables
with pionic quantum numbers, in such a way as to ensure that these identities
are realised. It should be noted that in this approach the pionic field has
no physical significance - it is merely an auxiliary field - and should not
be identified with the physical pion\cite{L92}.

At first nonleading order in ChPT, O(E$^4$), the effective action is only
completely determined once the effect of one-pion loops, generated by the
O(E$^2$) part of the action, is included.  The regularisation of the
divergence of each of these loops introduces ten arbitrary parameters at this
level.  The theory is not renormalisable but the approach is nevertheless
useful because higher order loops do not contribute to the O(E$^4$)
action\cite{W79}; i.e., the action at O(E$^4$) is completely specified once
the effect of the O(E$^2$) loops is taken into account and the 10 parameters
fixed by comparison with experimental data.

The pion (or, more generally, pseudoscalar) loops in ChPT are characteristic
of the approach and, indeed, are sometimes regarded as being the dominant
feature.  The expression for every physical observable receives a
contribution from such loops which depends on the mass of the particle in the
loop and which diverges in the chiral limit.  For example, in the case of the
electromagnetic pion charge radius the loop introduces a logarithmic
dependence and hence a logarithmic divergence in the chiral limit.  Herein we
are interested in estimating the importance of these loop contributions,
evaluated at real pseudoscalar masses, relative to what one might call the
``quark core'' contribution and our exemplary case will be the
electromagnetic charge radius of the pion.

The pseudoscalar mesons are the simplest bound states to study within the
coupled Dyson-Schwinger--Bethe-Salpeter equation approach to QCD, recent
reviews of which can be found in Refs.~\cite{H91,CDR93,RW94}.  In vector
exchange theories with DCSB, such as QCD, the Goldstone-boson character of
the pseudoscalar excitations can be understood through the existence of an
identity between the fermion Dyson-Schwinger equation (DSE) and the
pseudoscalar Bethe-Salpeter equation (BSE).  If, in the limit
\mbox{$P^2 \rightarrow 0$}, this BSE reduces to the DSE then DCSB necessarily
entails the existence of massless pseudoscalar excitations.  This result was
established for the rainbow-DSE--ladder-BSE pair in Ref.~\cite{DS79}.  (The
rainbow approximation in the DSE is equivalent to the ladder approximation in
the BSE.)  In this approach, the ``quark core'' of a meson is described by
the Bethe-Salpeter amplitude obtained as a solution of the pesudoscalar BSE;
the kernel of which, in practice, is often constructed as a generalised
ladder kernel: generalised in the sense that the quark propagators are
obtained as solutions of the rainbow DSE with a gluon propagator whose large
spacelike-$q^2$ behaviour is that given by the renormalisation group in QCD
and whose small spacelike-$q^2$ behaviour is modelled so as to ensure
confinement\cite{PCR89,AKM91,JM92,SC93}.

The coupled rainbow-DSE--ladder-BSE approach to modelling QCD is equivalent
to the Global Colour-symmetry Model (GCM), first described in
Ref.~\cite{CR85} and extended to the pseudoscalar sector in
Refs.~\cite{RCP88,RCSI93}.  The extension to other mesons and baryons is
reviewed in Ref.~\cite{C92}.  As discussed in Ref.~\cite{RCSI93}, this
approach provides a very good description of kinematic and dynamical
properties of pions. For example, using the quark core alone, it describes
$\pi$-$\pi$ partial wave amplitudes very well up to $600$~MeV; i.e., from
threshold up to the first resonance.  As indicated above, the parameters in
the GCM describe the quark-quark interaction at small spacelike-$q^2$.

The model represents QCD as an effective two-point theory, coloured quark
currents interact via dressed gluon exchange, and has the ``disadvantage''
that there is some model dependence in the results: for example, it predicts
values for the parameters that appear in the effective lagrangian of
ChPT\cite{RCSI93,HTV90}.  However, since these parameters have been fitted to
low energy phenomena in QCD, this provides a means of checking the model
assumptions in the GCM.  Thus the model dependence is also an advantage: the
assumptions on which the model is founded can be tested and this enables the
development of an understanding of hadronic phenomena in terms of quark and
gluon interactions.

Herein, in the context of the GCM, we calculate the electromagnetic charge
radius of the pion taking into account both the ``quark core'' and the
pion-loop rescattering correction.  The ``quark core'' contribution, obtained
in a generalised impulse approximation\cite{RCSI93,FT93,CDRrpi}, is regular in
the chiral limit whereas, as noted above, the pion-loop contribution is not.
Our aim is to estimate which is the dominant piece at the physical pion mass
and hence to infer whether the electromagnetic structure of the pion is
dominated by the ``quark core'' or a ``pseudoscalar meson-cloud''.

In Sec.~II we describe the calculation of \mbox{$(r_\pi^2)^{\rm GIA}$}, the
generalised impulse approximation ``quark core'' contribution to the charge
radius.  In Sec.~III we discuss the calculation of the pion loop contribution
to the charge radius.  We put these pieces together to obtain the net
$\pi\pi\gamma$ vertex and obtain an expression for the net charge form
factor in Sec.~IV.  Our results are presented and discussed in Sec.~V.  We
summarise and conclude in Sec.~VI.

\section{Generalised impulse approximation}
In Euclidean space, with metric \mbox{$\delta_{\mu\nu}={\rm diag}(1,1,1,1)$},
$\gamma_\mu$ hermitian and $m_u=m_d$, we define the generalised impulse
approximation contribution to the connected $\pi\pi\gamma$ vertex to
be:\cite{RCSI93,FT93,CDRrpi}
\begin{eqnarray}
\label{LFpi}
\lefteqn{\Lambda_\mu^{\rm GIA}(p,q)=} \\
& & \frac{N_c}{f_\pi^2}\,
\int\case{d^4k}{(2\pi)^4}\, {\rm tr}_D
\left[ \overline{\Gamma}_\pi(k_{0+};p)
        S(k_{++})i\Gamma_\mu(k_{++},k_{--})S(k_{--})
       \Gamma_\pi(k_{-0};q) S(k_{-+})\right]~.
\nonumber
\end{eqnarray}
In this expression the trace over colour and flavour indices has been carried
out leaving only the trace over Dirac indices and the definition
\begin{equation}
k_{\alpha\beta} = k + \case{\alpha}{2}p + \case{\beta}{2} q 
\end{equation}
has been used.  The dressed quark-photon vertex is denoted by
$\Gamma_\mu(p_1,p_2)$, the pion Bethe-Salpeter amplitude by $\Gamma_\pi(p;P)$
and the dressed quark propagator by $S(p)$.  This vertex is illustrated in
Fig.~\ref{figppA} and is what we mean by the ``quark core'' contribution to
the electromagnetic properties of the pion.

\subsection{Quark propagator}
We describe equation~(\ref{LFpi}) as ``generalised impulse approximation''
for a number of reasons.  First, the internal quark lines are described by
dressed quark propagators that are obtained by solving the quark
Dyson-Schwinger equation (DSE):
\begin{equation}
\label{QDSE}
 S^{-1}(p) = i \gamma\cdot p +   m
     + \case{4}{3} g^2 \int \case{d^4k}{(2\pi)^4} \gamma_\mu
         S(k) \Gamma_\nu^g (k,p) D_{\mu \nu}((p-k)^2).     \label{fullDSE}
\end{equation}
In this equation $D_{\mu\nu}(k)$ is the dressed gluon propagator and
$\Gamma_\nu^g(p_1,p_2)$ is the dressed quark-gluon vertex; each of these
quantities satisfies its own DSE.  The general solution of this equation has
the form
\begin{equation}
S(p) = - i \gamma\cdot p\, \sigma_V(p^2) + \sigma_S(p^2)~.
\end{equation}
A commonly used equivalent representation is to specify the inverse in terms
of two scalar functions:
\begin{equation}
\label{Sinv}
S^{-1}(p) = i \gamma\cdot p\, A(p^2) + m + B(p^2)
\end{equation}
where $m$ is the current quark mass.  Both forms will be used herein.  The
commonly used ``rainbow approximation'' DSE is obtained from
Eq.~(\ref{fullDSE}) by writing
\begin{equation}
\label{RainDSE}
\Gamma_\nu^g (k,p) = \gamma_\nu~.
\end{equation}

Equation~(\ref{fullDSE}) has been studied extensively using many model forms
for the dressed gluon propagator and quark-gluon vertex and some of this work
is summarised in Refs.~\cite{H91,CDR93,RW94}.  A good deal has been learnt
about the analytic properties of the quark propagator in the complex
plane\cite{WickR,BRW92} and this is important in the calculation of many
observables.  In connection with the electromagnetic charge radius of the
pion, however, it is sufficient to have a representation (or interpolation) of
the quark propagator that is valid on the real spacelike axis.  One such
simple form, which is a modification of that used in Ref.~\cite{PS79} and a
simplification of the form used in Ref.~\cite{RCSI93}, is
\begin{eqnarray}
A(p^2)& =& 1~, \label{PSA} \\
B(p^2) & = & \frac{\Lambda_1^3}{p^2 + \Lambda_2^2} \label{PSB}~,
\end{eqnarray}
where reasonable values of the parameters are 
\mbox{$\Lambda_1 \sim 0.5 $~GeV~$\sim \Lambda_2$}.  
We remark that the fact that \mbox{$B\neq 0$}, even when the quark bare mass is
zero, is a manifestation of DCSB and, in general, for the small quark bare
masses that are appropriate for the pion sector, $A(p^2)$ and $B(p^2)$ are
insensitive to $m$; a result that can be seen clearly in the model of
Ref.~\cite{BRW92}.  This means that in our parametrisation of the DSE results
$\Lambda_1$ and $\Lambda_2$ are also insensitive to $m$.  We emphasise that
these forms are only suitable for interpolation on the real spacelike axis.
We will discuss this Ansatz in more detail below [see Sec.~\ref{QP}].

\subsection{Pion Bethe-Salpeter amplitude}
\label{SecPBSA}
The generalised impulse approximation, in addition to using dressed quark
propagators, uses a pion Bethe-Salpeter amplitude that is obtained as the
solution of the homogeneous Bethe-Salpeter equation (BSE):
\begin{equation}
\label{piBSE}
\Gamma^{rs}_\pi(p;P) = 
\int\,\case{d^4k}{(2\pi)^4} \,K^{rs;tu}(k,p;P)\,
\left(S(k-\case{1}{2}P)\Gamma_\pi(k;P)S(k+\case{1}{2}P)\right)^{tu}
\end{equation}
where $P$ is the centre-of-mass momentum of the bound state, $p$ is the
relative momentum between the quarks in the bound state, and the
superscripts, which run from $1$-$4$, are associated with the Dirac structure
of the amplitude.  The kernel, $K^{rs;tu}(p,k;P)$, has no flavour structure
in the isospin symmetric limit $m_u=m_d$ and since
\mbox{$\Gamma_\pi(p;P)\propto I_C$} and \mbox{$S(p)\propto I_C$} then
\mbox{$K^{rs;tu}(p,k;P)\propto I_C$}, where $I_C$ is the identity in colour
space.  The ladder approximation BSE is obtained from Eq.~(\ref{piBSE}) by
writing
\begin{equation}
\label{LaddBSE}
K^{rs;tu}(k,p;P)
= \case{4}{3}g^2 D_{\mu\nu}(p-k) \left(\gamma_\mu\right)^{rt}
\left(\gamma_\mu\right)^{us}~.
\end{equation}
Here $g^2\,D_{\mu\nu}$ is a model dressed gluon propagator which, in Landau
gauge, has the form:
\begin{equation}
\frac{g^2}{4\pi}\,D_{\mu\nu}(k) = 
\left(\delta_{\mu\nu} - \frac{k_\mu k_\nu}{k^2}\right)\,D(k^2)~.
\end{equation}
The behaviour of $D(k^2)$ at large spacelike-$k^2$ is determined from the QCD
renormalisation group:
\begin{equation}
\label{Dklgq}
D(k^2) \; \stackrel{k^2 \rightarrow \infty}{\approx} \;
\frac{1}{k^2}\;\frac{\lambda \pi}{\ln\frac{k^2}{\Lambda_{\rm QCD}^2}}
\end{equation}
with $\lambda = 12/[33-2 N_f]$ and where $\Lambda_{\rm QCD}\approx 0.2$~GeV
is the renormalisation point invariant QCD mass scale.  The behaviour of
$D(k^2)$ at small and intermediate spacelike-$k^2$ is unknown and in the GCM
[or coupled DSE-BSE approach] a few parameters are used to specify a model
form; i.e., the behaviour of the quark-quark interaction in the infrared is
parametrised.  The results one obtains for physical quantities in this
approach are completely determined by the values of these parameters, as
emphasised in Ref.~\cite{RCSI93}.  The constraints and commonly used
parametrisations are discussed in detail in Refs.~\cite{H91,CDR93,RW94} and
one efficacious form has
\begin{equation}
D(q^2) = C\,\Lambda_{\rm QCD}^2\,\delta^4(q) 
        + \frac{\lambda \pi}
        {q^2\,\ln\left(\tau + \frac{q^2}{\Lambda_{\rm QCD}^2}\right)}
\end{equation}
with $C= (3\pi)^3$ and $\tau = 3$~\cite{PCR89}. 

The most general form of the pseudoscalar amplitude, allowed by Lorentz
covariance, which is odd under parity transformations is\cite{LS69}
\begin{equation}
\label{GGP}
\Gamma_\pi(p;P) = \gamma_5 \left\{\,E(p,P) 
   - i (\gamma\cdot p) \,(p\cdot P) \,F(p,P)
   - i (\gamma\cdot P) \,G(p,P) - [\gamma\cdot p ,\gamma\cdot P] \,H(p,P)
\right\}
\end{equation}
and since $\pi^0$ is even under charge-conjugation then $E$, $F$, $G$ and $H$
are even functions of $(p\cdot P)$ in this case.

The coupled rainbow-DSE--ladder-BSE system, Eq.~(\ref{fullDSE}) with
(\ref{RainDSE}) plus Eq.~(\ref{piBSE}) with (\ref{LaddBSE}), which is
equivalent to the GCM effective action approach to QCD, has been studied
extensively and provides an excellent description of light-light, heavy-light
and heavy-heavy meson systems\cite{PCR89,AKM91,JM92,SC93}. These studies show
that it is a good approximation to write
\begin{equation}
\label{approxGGP}
\Gamma_\pi(p;P) = \gamma_5 E(p,P)~.
\end{equation}

In the chiral limit; i.e., when the current quark mass, $m$, is zero, the
pseudoscalar ladder-BSE and quark rainbow-DSE are identical\cite{DS79} and
one has a massless excitation in the pseudoscalar channel with
\begin{equation}
E(p,P^2=0) \propto B(p^2)
\end{equation}
if  the quark DSE allows a dynamically generated mass function:
\mbox{$B(p^2)\neq 0$}; i.e, if there is DCSB.  This is the manner in which
Goldstone's theorem is realised in the GCM and coupled DSE-BSE approach.  In
this case the parametrisation of Eq.~(\ref{PSB}) is also a good representation
of the pion Bethe-Salpeter amplitude. 

The normalisation of the amplitude is fixed, as usual, by requiring that, for
\mbox{$P^2 = -m_\pi^2$}:
\begin{eqnarray}
\lefteqn{2 f_\pi^2 P_\mu = 
 \int\,\case{d^4k}{(2\pi)^4}\,\case{d^4q}{(2\pi)^4}\,{\rm tr}_D \left[ 
\overline{\Gamma}_\pi(k;P) 
        \frac{\partial K(k,q;P)}{\partial P_\mu} \Gamma_\pi(q;-P)
\right] + }
\label{piNorm}\\
& & 
N_c\,\int\,\case{d^4k}{(2\pi)^4}\,{\rm tr}_D \left[ 
\overline{\Gamma}_\pi(k;P) S(k_{0-}) \Gamma_\pi(k;-P) 
        \frac{\partial S(k_{0+})}{\partial P_\mu}
+\overline{\Gamma}_\pi(k;P) \frac{\partial S(k_{0-})}{\partial P_\mu} 
        \Gamma_\pi(k;-P) S(k_{0+})
\right]. \nonumber
\end{eqnarray}
If the Bethe-Salpeter interaction kernel, $ K(k,p;P)$, is independent of the
centre-of-mass momentum, $P$, which is the case in ladder approximation, for
example, then this reduces to
\begin{eqnarray}
\label{LAN}
\lefteqn{2 f_\pi^2 P_\mu = }\\
& & 
N_c\,\int\,\case{d^4k}{(2\pi)^4}\,{\rm tr}_D \left[ 
\overline{\Gamma}_\pi(k;P) S(k_{0-}) \Gamma_\pi(k;-P) 
        \frac{\partial\,S(k_{0+})}{\partial P_\mu}
+\overline{\Gamma}_\pi(k;P) \frac{\partial\,S(k_{0-}) }{\partial P_\mu}
        \Gamma_\pi(k;-P) S(k_{0+})
\right]. \nonumber
\end{eqnarray}

\subsection{Quark-photon vertex}
\label{QP}
The generalised impulse approximation contribution also involves the dressed
quark-photon vertex $\Gamma_\mu(p_1,p_2)$ which satisfies its own integral
equation.  Solving this equation is a difficult problem which has only
recently begun to be addressed\cite{MF93}.  However, it is clear that if the
quark propagator has momentum dependent dressing then $\Gamma_\mu(p_1,p_2)$
cannot be the bare vertex since the Ward Identity
\begin{equation}
\label{WTI}
(p_1 - p_2)_\mu i \Gamma_\mu(p_1,p_2) = S^{-1}(p_1) - S^{-1}(p_2)
\end{equation}
would not then be satisfied and hence the electromagnetic current of the pion
would not be conserved.

It is possible to learn much about this vertex without actually solving its
integral equation.  Indeed, it is argued in
Refs.~\cite{CPcoll,QEDCJB,BPR92,BR93} that a ``physically reasonable'' Ansatz
for this vertex must take the form:
\begin{equation}
\Gamma_\mu(p,k) = \Gamma_\mu^{\rm BC}(p,k) + \Gamma_\mu^{\rm T}(p,k)
\end{equation}
where\cite{BC80}
\begin{eqnarray}
\Gamma_{\mu}^{\rm BC}(p,k) &  = &
\frac{\left[A(p^2) +A(k^2)\right]}{2}\;\gamma_{\mu} \nonumber \\
 & + &
\frac{(p+k)_{\mu}}{p^2 -k^2}\left\{ \left[ A(p^2)-A(k^2)\right]
                 \frac{\left[ \gamma\cdot p + \gamma\cdot k\right]}{2}
- i\left[ B(p^2) - B(k^2)\right]\right\}  \label{VBC}
\end{eqnarray}
and
\begin{equation}
(p-k)_\mu\,\Gamma_\mu^{\rm T}(p,k) = 0 \;\;{\rm with} \;\;
\Gamma_\mu^{\rm T}(p,p) = 0.
\end{equation}
In the bare quark propagator limit, $A=1$ and $B=$~constant,
\mbox{$\Gamma_\mu^{\rm T}=0$}. This Ansatz is ``physically reasonable''
because it satisfies the criteria a)-d) discussed in Ref.~\cite{BR93}: a) it
satisfies the Ward-Takahashi identity; b) it is free of kinematic
singularities; c) it reduces to the bare vertex in the free field limit; d)
it has the same transformation properties as the bare vertex under charge
conjugation and Lorentz transformations.  The
\mbox{$\Gamma_\mu^{\rm BC}(p,k)$} piece is completely determined by the quark
propagator but the remaining piece, \mbox{$\Gamma_\mu^{\rm T}(k,p)$}, is
undetermined.  

It is in this additional transverse piece that photon--vector-meson mixing
contributions will appear; i.e., vector meson dominance contributions are
confined to \mbox{$\Gamma_\mu^{\rm T}(p,k)$}.  This is especially clear in the
Nambu--Jona-Lasinio (NJL) model where
\mbox{$\Gamma_\mu^{\rm BC}(p,k) = \gamma_\mu$}\cite{BHS88,BM88}.  

We will define our generalised impulse approximation as Eq.~(\ref{LFpi}) with
\begin{equation}
\label{GTZ}
\Gamma_\mu^{\rm T}(p,k) \equiv 0~.
\end{equation}
This allows for a very good description of the electromagnetic form factor in
the spacelike region; i.e., away from resonance contributions\cite{CDRrpi}.
We note that criterion c) is important because, in association with a
solution of the DSE for the quark propagator which exhibits the correct
perturbative QCD leading-log behaviour of S(p), which is guaranteed when the
gluon propagator in Eq.~(\ref{QDSE}) has the ultraviolet form given in
Eq.~(\ref{Dklgq}), it ensures that the vertex Ansatz will yield the same
large-$Q^2$ behaviour for the pion form factor as is predicted by
QCD\cite{CDRrpi}.

With a quark-photon vertex that satisfies the above constraints and using the
identities:
\begin{eqnarray}
\label{Ia}
S(-k)^T& =& C^T\,S(k)\,C~, \\
\label{Ib}
\overline{\Gamma}_\pi^T(-k;-p) & = & - C^T\,\Gamma_\pi(k;p)\,C ~,\\
\label{Ic}
\Gamma_\mu^T(-k,-p) & = & - C^T\,\Gamma_\mu(p,k)\,C~,
\end{eqnarray}
where \mbox{$C=\gamma_2\gamma_4$} is the charge conjugation matrix, it
is easy to establish that
\begin{equation}
\label{GIACC}
(p+q)_\mu \Lambda_\mu^{\rm GIA}(p,q) = 0~;
\end{equation}
i.e., the $\pi$-current is conserved.  Hence one can write
\begin{equation}
\label{LCCF}
\Lambda_\mu^{\rm GIA}(p,q) = T_\mu(p,q)\,\hat{F}_\pi^{\rm GIA}(p^2,p\cdot q,q^2)
\end{equation}
where
\begin{equation}
T_\mu(p,q) = 2\, \frac{p_\mu q\cdot(p+q) - q_\mu p\cdot(p+q)}{(p+q)^2}~.
\end{equation}
In the case of elastic scattering \mbox{$p^2 = q^2$} and therefore
Eq.~(\ref{LCCF}) reduces to
\begin{equation}
\Lambda_\mu^{\rm GIA}(p,q) = 
(p-q)_\mu F_\pi^{\rm GIA}((p+q)^2,p^2)~.
\end{equation}

Using Eqs.~(\ref{Ia}-\ref{Ic}) one can also show that 
\mbox{$\Lambda_\mu^{\rm GIA}(P,-P)$} is 
\begin{eqnarray}
\label{GIAnorm}
2 P_\mu\,F^{\rm GIA}(0,P^2)  &= & \nonumber
\frac{N_c}{f_\pi^2} \int\case{d^4 k}{(2\pi)^4} 
{\rm tr}_D \left[ 
\overline{\Gamma}_\pi(k;P) S(k_{-0}) \Gamma_\pi(k;-P) 
        \frac{\partial}{\partial P_\mu}\,S(k_{+0}) \right. \\
& & \;\;\;\;\;\;\;\;\;\;\;\;\;\;\;\;\;\;\;\;
\left. +\overline{\Gamma}_\pi(k;P) \frac{\partial}{\partial P_\mu}\,S(k_{-0}) 
        \Gamma_\pi(k;-P) S(k_{+0})
\right]~. 
\end{eqnarray}
Comparing this with Eqs.~(\ref{piNorm}) and (\ref{LAN}) one observes that in
generalised impulse approximation \mbox{$F^{\rm GIA}(0,P^2)=1$} only if the
Bethe-Salpeter kernel is independent of $P$; i.e., generalised impulse
approximation combined with a $P$-independent Bethe-Salpeter kernel provides
a consistent approximation scheme.  In this case one has, in the chiral limit
\mbox{$P^2 = -m_\pi^2 = 0$}:
\begin{eqnarray}
\lefteqn{f_{\pi}^2  = }\label{Fpi}\\
& & \frac{N_c}{8\pi^2}\int_{0}^\infty\,ds\,s\,\Gamma_\pi(s)^2\,
\left(  \sigma_{V}^2 - 
2 \left[\sigma_S\sigma_S' + s \sigma_{V}\sigma_{V}'\right]
- s \left[\sigma_S\sigma_S''- \left(\sigma_S'\right)^2\right]
- s^2 \left[\sigma_V\sigma_V''- \left(\sigma_V'\right)^2\right]\right)~.
\nonumber
\end{eqnarray}
This illustrates that Eq.~(\ref{LFpi}) is regular in the chiral limit and only
weakly dependent on $m_\pi$. 

We are now in a position to consider the Ansatz for the quark propagator,
Eqs.~(\ref{PSA}) and (\ref{PSB}), in more detail.  Following
Ref.~\cite{RCSI93}, this Ansatz for the fermion propagator can be used to fit
the characteristic parameters of the $\pi$-$\pi$ sector with the result that,
with
\begin{eqnarray}
\label{Lvalues}
\Lambda_1 = 0.527\;{\rm GeV} &\;\;{\rm and} \;\; 
& \Lambda_2 = 0.573 \;{\rm GeV},  
\end{eqnarray}
one obtains 
\begin{equation}
\label{fpirpi}
\begin{array}{cc}
f_\pi = 0.093\,{\rm GeV}\;(0.093), \;\; &  r_\pi = 0.53\,{\rm fm}\;(0.66)
\end{array}
\end{equation}
and the following values for the dimensionless scattering lengths:
\begin{equation}
\begin{array}{lll}
 a_{0}^0 =   0.16 \;(0.20),\;& a_{0}^2 = -0.047\;(-0.037),\;&  \\
a_{1}^1 =  0.029\;(0.038),\;  & a_{2}^0 =0.0014\;(0.0017),\;&
 a_{2}^2 = -0.00037~,
\end{array}
\end{equation}
where the experimental values, when known, are given in parentheses.  These
results were obtained using the approximation of Eq.~(\ref{approxGGP}) and
$f_\pi$ was calculated using Eq.~(\ref{Fpi}).  The formulae for the remaining
quantites are all given in Ref.~\cite{RCSI93} and, apart from
\mbox{$(r_\pi^2)^{\rm GIA}$} in Eq.~(\ref{AppRpi}), will not be
reproduced\cite{EQnos}, however, all of the information needed to derive them
is contained herein. The results are simply presented so as to demonstrate
the efficacy and utility of this simple Ansatz and to fix reasonable values
of the parameters.

\subsection{Approximate form of $F^{\rm GIA}(p^2,p\cdot q, q^2)$}
All of the elements of the calculation of \mbox{$F^{\rm GIA}(p^2,p\cdot q,
q^2)$} have now been described.  The actual calculation involves a
three-dimensional integral that can be evaluated numerically without
difficulty for arbitrarily large spacelike momentum transfer provided a
sensible DSE-BSE framework is used to describe the quark-gluon substructure
of the pion. This is the subject of Ref.~\cite{CDRrpi}.  (The fact that
Eqs.~(\ref{PSA}) and (\ref{PSB}) should only be used for interpolation is
important in this context.)  Herein we are interested in the charge radius as
it is affected by this parton substructure and by meson self-dressing.  The
detailed structure of \mbox{$F^{\rm GIA}(p^2,p\cdot q, q^2)$} is not
necessary for this calculation.

In general \mbox{$F^{\rm GIA}(p^2,p\cdot q,q^2)$} depends on all three of its
arguments and, as will become clear below, this dependence will be sampled by
the pion loop contribution to the charge radius.  However, our calculation is
greatly simplified if we make the approximation that
\begin{equation}
\label{monofit}
F^{\rm GIA}(p^2,p\cdot q, q^2) \; = \; F^{\rm GIA}(Q^2),
\end{equation}
where $Q^2 = (p+q)^2$, for then this piece appears only as a multiplicative
factor in the pion loop contribution to the charge radius.  This
approximation is expected to lead to an overestimate of the importance of the
pion loop to the charge radius since any off-shell effects in \mbox{$F^{\rm
GIA}(p^2,p\cdot q, q^2)$} will lead to additional damping in the pion-loop
integral, which is over spacelike momenta.

Subject to this approximation then, for our purposes, all we need to know is
\begin{equation}
(r_\pi^2)^{\rm GIA}= - 6 \left. 
\frac{d}{dQ^2}\ln\,F^{\rm GIA}(Q^2)\right|_{Q^2=0}
\end{equation}
which has been presented in Ref.~\cite{RCSI93} and which we reproduce in
Eq.~(\ref{AppRpi}) of the appendix.  As we have seen, the simple
interpolating form for the quark propagator that we are using herein yields
the charge radius in Eq.~(\ref{fpirpi}) while the more sophisticated analysis
of Ref.~\cite{RCSI93}, using a somewhat different form of the fermion
propagator, obtained the result \mbox{$r_\pi = 0.59$ fm}.  Both of these
calculations are to be compared with the experimental value of \mbox{$r_\pi =
0.66$ fm}~\cite{SRA86}.  This illustrates what we believe to be a general
result: that the generalised impulse approximation consistently understimates
the charge radius by $< 15$\% leaving room for only small contributions
from meson-dressing effects such as the pion loop and $\rho^0$-photon mixing.

\section{Pion Loop Contribution}
We approximate the pion loop contribution to the connected $\pi\pi\gamma$
vertex as
\begin{eqnarray}
\label{PionLoop}
\lefteqn{\Lambda_\mu^{ij\,{\rm Loop}}(p,q)=} \\
& & -\case{1}{2}\,\epsilon^{3kl}\,\int\case{d^4k}{(2\pi)^4}\,
T^{ijkl}(p+q,p-k,k-q) \, D_\pi(k)\, D_\pi(p+q-k)\,
\Lambda_\mu^{\rm GIA}(k,p+q-k)\nonumber
\end{eqnarray}
where \mbox{$D_\pi(k)$} is the pion propagator,
\mbox{$\Lambda_\mu^{\rm GIA}(p,q)$} is given in Eq.~(\ref{LFpi})  and 
the $\pi$-$\pi$ scattering kernel is
\mbox{$T^{ijkl}(p_1+p_2,p_1+p_3,p_1+p_4)$}, which can be written in terms of a
single scalar function: 
\begin{eqnarray}
\label{Tijkl}
\lefteqn{T^{ijkl}(p_1+p_2,p_1+p_3,p_1+p_4)  = }\\
& & \delta^{ij}\delta^{kl}\,A(p_1,p_2;p_3,p_4) +
\delta^{ik}\delta^{jl}\,A(p_1,p_3;p_2,p_4) +
\delta^{il}\delta^{jk}\,A(p_1,p_4;p_3,p_2) \nonumber
\end{eqnarray}
with
\begin{equation}
A(p_1,p_2;p_3,p_4) = A(p_2,p_1;p_3,p_4) =
A(p_1,p_2;p_4,p_3) = A(p_2,p_1;p_4,p_3)
\end{equation}
from which all of the relations of crossing symmetry follow.  We also note
that the result
\begin{equation}
(p+q)_\mu \Lambda_\mu^{\rm Loop}(p,q) = 0~ 
\end{equation}
follows from Eq.~(\ref{GIACC}) and as a consequence one may write
\begin{equation}
\Lambda_\mu^{ij\,{\rm Loop}}(p,q) = \epsilon^{3ij}\,
T_\mu(p,q)\,\hat{F}_\pi^{\rm Loop}(p^2,p\cdot q,q^2)
\end{equation}
which, in the case of elastic scattering \mbox{$[p^2 = q^2]$}, reduces to
\begin{equation}
\Lambda_\mu^{ij\,{\rm Loop}}(p,q) = \epsilon^{3ij}\,
( p-q)_\mu F_\pi^{\rm Loop}((p+q)^2,p^2)~.
\end{equation}
Equation~(\ref{PionLoop}) is illustrated in Fig.~\ref{figpiloop}.  This is
the analogue in the GCM [and coupled DSE-BSE approach] of the one pion loop
diagram calculated in ChPT.

\subsection{$\pi$-$\pi$ scattering amplitude}
The $\pi$-$\pi$ scattering amplitude is an important part of
Eq.~(\ref{PionLoop}). The form of \mbox{$A(p_1,p_2;p_3,p_4)$} near threshold
in the GCM was studied in detail in Ref.~\cite{RCSI93}.  Therein it was shown
to reproduce the Weinberg term\cite{W66} at O(E$^2$) and to possess
additional structure at O(E$^4$) that provides for a better description of
the scattering lengths and partial wave amplitudes in $\pi$-$\pi$ scattering:
\begin{eqnarray}
\label{Astu}
A(s,t,u) & = & 
{{{m_{\pi}^2} + 2\,s - t - u}\over {3\,{f_{\pi}^{2}}}}\\
& + & \frac{4 N_c}{3 f_{\pi}^{4}}  
\left[ K_1 \left( -12\,{m_{\pi}^4} +
6\,{m_{\pi}^2}\,(s + t +u) + 2\,{s^2} - {t^2}- {u^2}  
- 2(\,s\,t +\,s\,u +\,t\,u ) \right)\right. \nonumber \\
& & + K_2 \left.
\left( -2\,{m_{\pi}^2} + s \right) \,\left( -2\,{m_{\pi}^2} + t + u \right)
+ K_3
\left(-2\,{m_{\pi}^4} + {m_{\pi}^2}\,(s +t+u)  -t\,u\right)\right]~, \nonumber
\end{eqnarray}
where $s=-(p_1+p_2)^2$, $t=-(p_1+p_3)^2$ and $u=-(p_1+p_4)^2$ are the usual
Mandelstam variables and in the GCM~\cite{RCSI93}:
\mbox{$K_1= 0.000508$}, \mbox{$K_2= 0.00930$} and \mbox{$K_3= 0.00101$}.

In the present study, however, an expansion in powers of momenta about
threshold is inadequate since the integration in Eq.~(\ref{PionLoop}) samples
momenta well away from threshold.  A broad ranging discussion of the
constraints that analyticity and unitarity place on the asymptotic behaviour
of the scattering amplitude can be found in Refs.~\cite{MJ64}; one result
being that the forward scattering amplitude, \mbox{$A(s,0,4m_\pi^2-s)$},
cannot fall faster than \mbox{$s^{-2}$}.  

The dominant contribution to \mbox{$A(p_1,p_2;p_3,p_4)$} in the GCM is
obtained from Eqs.~(A.14) and (A.16b) in Ref.~\cite{RCP88}: 
\begin{equation}
\label{ADs}
\int\,d^4x\,d^4y\,\frac{f_\pi^2}{2}{\rm tr}
\left[\partial_\mu U(x)\,\partial_\mu U(y)^{\dagger}\right] \hat{f}(x-y)
\end{equation}
where, as usual, \mbox{$U(x)=
\exp(i\bbox{\pi}\cdot\bbox{\tau}/f_\pi)$} and 
\begin{equation}
\hat{f}(x)= \int\case{d^4k}{(2\pi)^4}\, {\rm e}^{i k\cdot x} \,f(k^2)
\end{equation}
with
\begin{equation}
\label{fp}
f_\pi^2\,p^2\,f(p) = 2 N_c \int\case{d^4k}{(2\pi)^4}\, 
\Gamma_\pi(k^2)^2\,K(k_+)\,K(k_-)\, 
\left( k^2 A_-^2 + k\cdot p A_- A_+ + \case{1}{4}p^2\,A_+^2 \right)~,
\end{equation}
\mbox{$[k_\pm = k \pm \case{1}{2}p]$}, \mbox{$[A_\pm = A(k_+) \pm A(k_-)]$}
and where
\begin{equation}
K(p) = \frac{1}{p^2 A(p^2)^2 + B(p^2)^2}~.
\end{equation}
[This expression has been generalised here to allow for momentum dependence
of $A(p^2)$.]  In Eq.~(\ref{fp}), the right-hand-side contributes the leading
term in the expression for $f_\pi^2$ in the chiral limit: the other terms
that contribute to Eq.~(\ref{Fpi}) are obtained from Eqs.~(A.16a) and (A.16c)
in Ref.~\cite{RCP88} and in the point-meson [or Nambu--Jona-Lasinio model]
limit of the GCM, defined such that \mbox{$A(p^2)=1$} and \mbox{$B(p^2)= {\rm
constant}$}, these terms vanish.  Taking this into account we renormalise
Eq.~(\ref{fp}) and define
\begin{equation}
\label{fpN}
p^2\,f(p) = \frac{1}{\cal N'} \int\,\frac{d^4k}{2\pi^2}\, 
\Gamma_\pi(k^2)^2\,K(k_+)\,K(k_-)\, 
\left( k^2 A_-^2 + k\cdot p A_- A_+ +\case{1}{4}p^2\,A_+^2 \right)
\end{equation}
with ${\cal N}'$ chosen so that $f(p^2=0)=1$.  

The contribution of Eq.~(\ref{ADs}) to \mbox{$A(p_1,p_2;p_3,p_4)$} at tree
level is easily found to be
\begin{eqnarray}
\label{AIW} 
A^{\rm LO}(p_1,p_2;p_3,p_4) & = &
\frac{1}{3f_\pi^2} \left(
3 \,(p_1+p_2)^2 f(p_1+p_2) - \sum_{i=1}^{4} p_i^2 f(p_i^2) \right)~.
\end{eqnarray}
We refer to this expression as the ``leading-order'' scattering amplitude
because it provides the leading contribution at large $s$; i.e., it is the
contribution to the scattering amplitude in the GCM which falls least rapidly
with increasing $s$.  We note that neglecting the momentum dependence of
$f(p)$; i.e., at lowest order in a momentum expansion, Eq.~(\ref{AIW})
yields, upon continuation to Minkowski space,
\begin{eqnarray}
A^{\rm LO}(p_1,p_2;p_3,p_4) & = &
\frac{2s - t - u}{3f_\pi^2} 
\end{eqnarray}
which reproduces the result of Ref.~\cite{W66} in the chiral limit,
\mbox{$m_\pi\rightarrow 0$}.  

We would like an accurate estimate of this amplitude in the GCM.  It is not
difficult to evaluate the integral numerically using the results of DSE
studies, as represented by Eqs.~(\ref{PSA}), (\ref{PSB}) and (\ref{Lvalues}),
but it is more instructive to distill these ingredients and obtain an
analytic expression that manifests the important features.  To do this we use
Eqs.~(\ref{PSA}), (\ref{PSB}) and (\ref{approxGGP}) but, in addition, we make
a further simplifying assumption, based on the ``constituent mass'' concept
described in Ref.~\cite{PCR88}.  The basic observation in this connection is
that the integration in Eq.~(\ref{fp}) is weighted, and cut off, by the
factor \mbox{$k^2\,\Gamma_\pi(k^2)^2$} that arises because of the quark-gluon
substructure of the pion in the GCM.  This manifestation of the quark core
of the meson is a general feature of the coupled DSE-BSE approach and ensures
that even meson loop integrals are intrinsically finite.  This factor
concentrates the integration domain around \mbox{$s=\Lambda_2^2$}, using
Eq.~(\ref{approxGGP}), and hence it is a reasonable approximation to replace
\mbox{$B(p^2)$} in \mbox{$K(p^2)$} by an ``effective constituent quark
mass'':
\begin{equation}
\label{Kapprox}
K(p^2) = \frac{1}{p^2 + M_c^2}
\end{equation}
where a first estimate of $M_c$ is obtained by evaluating $B(p^2)$ at 
\mbox{$p^2 = \Lambda_2^2$}, which is the maximum of
\mbox{$k^2\,\Gamma_\pi(k^2)^2$}: 
\begin{equation} 
\label{MC}
M_c \approx B(\Lambda_2^2) = \frac{\Lambda_1^3}{2\,\Lambda_2^2}~.
\end{equation}
Using the values of $\Lambda_1$ and $\Lambda_2$ in Eq.~(\ref{Lvalues}) this
yields \mbox{$M_c \sim 0.22$ GeV}.

A more sophisticated estimate of $M_c$ may be obtained by requiring that
using Eq.~(\ref{Kapprox}) in Eq.~(\ref{Fpi}) should yield the same result for
$f_\pi$ as the exact calculation. Using the approximations of
Eqs.(\ref{PSA}), (\ref{PSB}), (\ref{approxGGP}) one can rewrite
Eq.~(\ref{Fpi}) as
\begin{eqnarray}
f_\pi^2 & = &\frac{N_c}{8\pi^2}\int_{0}^\infty\,ds\,s\,K(s)^2\,B(s) 
\left(B(s) - \case{1}{2}B'(s)\right)~,
\end{eqnarray}
which, not unexpectedly, is the result presented in Ref.~\cite{PS79}.
Substituting Eq.~(\ref{Kapprox}) for $K$ in this expression one obtains
\begin{eqnarray}
f_\pi^2 
& = & \frac{N_c\Lambda_1^6}{16\pi^2} \,
\frac{\left(-7\,\Lambda_2^4 + 20\,\Lambda_2^2\,M_c^2 - 13\,M_c^4
        +\ln\frac{M_c^2}{\Lambda_2^2}
        \left[-4\,\Lambda_2^4 + 4 \,\Lambda_2^2\,M_c^2 + 6\,M_c^4\right]\right)}
        {(M_c^2-\Lambda_2^2)^4}
\end{eqnarray}
and requiring that this yield \mbox{$f_\pi = 0.093$ GeV} gives
\begin{equation}
M_c = 0.31 \;{\rm GeV}~,
\end{equation}
which shows that Eq.~(\ref{MC}) provides a reasonable estimate.

Using Eqs.~(\ref{PSA}), (\ref{PSB}), (\ref{approxGGP}) and (\ref{Kapprox}) we
obtain
\begin{equation}
\label{fpapprox}
p^2 f(p) = \frac{2}{\cal N}\,
\int_0^\infty \,du\, \frac{1}{(u+\Lambda_2^2)^2}\,
\left(1 - \frac{\sqrt{\left(u + \case{1}{4}p^2 + M_c^2\right)^2 - u p^2 }}
        {u + \case{1}{4}p^2 + M_c^2}\right)
\end{equation}
with
\begin{equation}
\label{Norm}
{\cal N} = \frac{ (M_c^2 + \Lambda_2^2)\,\ln\frac{M_c^2}{\Lambda_2^2}
                - 2\,(M_c^2 -\Lambda_2^2)}
                {(M_c^2 - \Lambda_2^2)^3}~.
\end{equation}
It is not difficult to establish that, at large-$p^2$, 
\begin{equation}
f(p) \approx \frac{1}{\cal N} \frac{8}{p^4} 
\ln\frac{p^2}{\Lambda_2^2} 
\end{equation}
which gives \mbox{$A^{\rm LO}(s,t,u)$} consistent with the bounds of
Refs.~\cite{MJ64}.  We emphasise that $\Lambda_2$ is not an arbitrary
parameter introduced to regularise the integral but, rather, it is a direct
measure of the quark-gluon substructure in the model arising as it does
through solving the Bethe-Salpeter equation; i.e., it is a calculated quantity in
the GCM.  We note also that $\Lambda_1$ has not disappeared but is contained
implicitly in $M_c$.

\section{Net $\pi\pi\gamma$ Vertex}
The net $\pi\pi\gamma$ vertex is obtained by adding Eq.~(\ref{PionLoop})
directly to Eq.~(\ref{LFpi}):
\begin{eqnarray}
\Lambda_\mu^{ij}(p,q) & = &\epsilon^{3ij} \Lambda_\mu^{\rm GIA}(p,q)
+\Lambda_\mu^{ij\,{\rm Loop}}(p,q) \\
& \equiv & \epsilon^{3ij}\,T_\mu(p,q)\,\hat{F}_\pi(p^2,p\cdot q,q^2) 
\end{eqnarray}
with
\begin{equation}
\label{PFF}
  \hat{F}_\pi(p^2,p\cdot q,q^2)  = 
\hat{F}_\pi^{\rm GIA}(p^2,p\cdot q,q^2) + 
\hat{F}_\pi^{\rm Loop}(p^2,p\cdot q,q^2)~.
\end{equation}
This does not lead to double-counting because the pion
final-state-interactions cannot be recast into an additional dressing of the
quark-photon vertex; unlike the $\rho$-meson contribution that is contained
in \mbox{$\Gamma^T(p,k)$} and which we neglected.  Adding
Eq.~(\ref{PionLoop}) is thus a correction beyond generalised impulse
approximation and as such leads to a renormalisation of the pion field in our
approach, which is simply a reordering of the summation of diagrams in a well
defined and systematic manner as prescribed by the bosonisation of the GCM.

To clarify this statement we remark that going beyond generalised impulse
approximation in the calculation of the electromagnetic pion form factor
builds additional structure into the pion. In the framework of the GCM, the
``bare pion''; i.e., the one whose Bethe-Salpeter amplitude appears in
Eq.~(\ref{LFpi}), is a ladder $q$-$\overline{q}$ bound state: it represents
the ``quark core'' of the pion.  As shown in Ref.~\cite{RCSI93}, this core
provides a very good description of kinematic and dynamical properties of the
pion away from resonance contributions.  Adding the pion loop contribution is
the first step in allowing this ladder $q$-$\overline{q}$ bound state to
dress itself with a cloud of ladder mesonic bound states.  In the framework
of the BSE, Eq.~(\ref{piBSE}), this corresponds to going beyond ladder
approximation.  Indeed, this modification could be built into the pion BSE
and would lead to a dependence on the centre-of-mass momentum, $P$, in the
kernel of Eq.~(\ref{piBSE}) thus changing the normalisation obtained from
Eq.~(\ref{piNorm}).  This illustrates the nature of the GCM approach.  The
``tree-level'' fields in the effective action represent ladder
$q$-$\overline{q}$ bound states and the remaining interaction terms provide
for meson self-dressing, in addition to meson-meson interactions, which
extends the nature of the structure of the bound state beyond that of ladder
approximation.

The pion propagator in  Eq.~(\ref{PionLoop}) has the form
\begin{equation}
D_\pi(k) = \frac{1}{k^2 + \Sigma_\pi^2(k^2)}
\end{equation}
where \mbox{$\Sigma_\pi^2(k^2)$} is the pion self energy which can be
obtained as the solution of a DSE or using the eigen-value procedure of
Ref.~\cite{C92}.  Herein we follow Ref.~\cite{HRM92} and use the
approximation
\begin{equation}
\label{PiProp}
D_\pi(k) = \frac{1}{k^2 + (m_\pi^{\rm L})^2}~,
\end{equation}
where \mbox{$(m_\pi^{\rm L})^2 - \Sigma_{\pi}^2(-(m_\pi^{\rm L})^2)=0$}.  We
have labelled the mass of the pion in the loop \mbox{$m_\pi^{\rm L}$} in
order to artificially distinguish it from the mass of the external pions; a
convention we adopt simply so as to make explicit the dependence of the
charge radius on the mass of the loop particle.  Equation~(\ref{PiProp}) is
expected to be a good approximation because the Bethe-Salpeter amplitudes in
Eqs.~(\ref{LFpi}) and (\ref{PionLoop}) ensure that the dominant integration
domain is at small $k^2$ where the pion propagator should not be much
modified from its on-shell form.  Indeed, the dominant dressing contributions
for the virtual pion propagator would come from a 3$\pi$ intermediate state
and this was shown to be negligible in Ref.~\cite{HRM92}.

Using Eq.~(\ref{PiProp}) in Eq.~(\ref{PFF}) we obtain the following general
expression for the one-pion-loop corrected approximation to the
electromagnetic pion form factor:
\begin{eqnarray}
\label{PiRpifull}
\lefteqn{
\hat{F}_\pi(p^2,p\cdot q,q^2)  =  
\hat{F}_\pi^{\rm GIA}(p^2,p\cdot q,q^2) }\\
&  & - \frac{1}{2}\int\frac{d^4k}{(2\pi)^4}\, \left[
\frac{q\cdot(p+q)\,k\cdot(p+q) - q\cdot k\,(p+q)^2}
        {p^2q^2 - (p\cdot q)^2} \,\times \right. \nonumber\\
& & \left.
\frac{A(p,-k;q,k-p-q) - A(p,k-p-q;q,-k)}
        {(k^2+(m_\pi^{\rm L})^2)\,((-k+p+q)^2+(m_\pi^{\rm L})^2)}\,
\hat{F}_\pi^{\rm GIA}(k^2,k\cdot (-k+p+q),(-k+p+q)^2) \right].
\nonumber
\end{eqnarray}
As we have mentioned above and make explicit below, the pion loop
contribution is given by a finite integral because the ``bare meson'' in the
loop already has an internal structure: its $q$-$\overline{q}$ core as given
by the BSE, Sec.~\ref{SecPBSA}.

\section{Pion Charge Radius}
Equation~(\ref{PiRpifull}) is simplified by the approximations we have
described above.  Using Eq.~(\ref{monofit}), the ``leading order'' scattering
amplitude, Eqs.~(\ref{AIW}) and (\ref{fpapprox}), shifting the integration
variable:
\mbox{$[k\rightarrow \, k+q]$}, writing 
\begin{eqnarray}
P= p-q &\;\; {\rm and} \;\; & Q = p+q~,
\end{eqnarray}
introducing hyperspherical polar coordinates and placing the external pions
on the chiral-limit mass-shell, \mbox{$[P^2 + Q^2 = 0]$}, we
obtain the following expression\cite{fn}:
\begin{eqnarray}
\label{PiRpi}
\hat{F}_\pi(Q^2)  & =  & \hat{F}_\pi^{\rm GIA}(Q^2) 
\left\{ 1 +\frac{1}{16\pi^3\,f_\pi^2} \left[\int_0^\infty\,dk\,k^3\,
        \int_{-1}^{1}dy\, \left(1+2iy\frac{k}{Q}\right) \right. \right.\\
& & \left. \times\,
        \frac{(k^2 - 2 iQ k y - Q^2) f(k^2 - 2 iQ k y - Q^2) - k^2 f(k^2)}
        {k\, Q\, (k^2 - i k Q y + (m_\pi^{\rm L})^2 )} 
\right. \nonumber\\
& & \left.\left. \times
\ln\left|
\frac{k^2 - i k Q y + (m_\pi^{\rm L})^2 + k Q \sqrt{1-y^2}}
        {k^2 - i k Q y + (m_\pi^{\rm L})^2 - k Q \sqrt{1-y^2}}\right|
\right]\right\}\nonumber
\end{eqnarray}

Equation (\ref{PiRpi}) can be used to calculate the one-pion-loop corrected
value of the pion decay constant, which we will denote
\mbox{$\hat{f}_\pi$},
\begin{equation}
\label{FpiRen}
\hat{f}_\pi^2 = f_\pi^2 \, \hat{F}_\pi(0).
\end{equation}
A little algebra yields
\begin{equation}
\hat{f}_\pi^2 - f_\pi^2  = 
               \frac{ 2}{\pi^2 }
               \int_0^\infty \,du\, \frac{u^2}{[4u + (m_\pi^{\rm L})^2]^2}
                                   \,g(u) 
\end{equation}
where, with $f(x)$ given in Eq.~(\ref{fpapprox}),
\begin{eqnarray}
g(u) & = & f(4\,u) + u \,\frac{d}{du} f(4\,u) \\
& = &\frac{1}{\cal N}\,\int_0^\infty dv 
\frac{v(v+M_c^2-u)}
{(v+\Lambda_2^2)^2 (v+M_c^2+u)^2\sqrt{(v+M_c^2+u)^2-4uv}} 
\label{guint}
\end{eqnarray}
and ${\cal N}$ is given in Eq.~(\ref{Norm}).  This integral can be evaluated
and the result is given in Eq.~(\ref{guExp}) of the appendix.

The result in Eq.~(\ref{fpirpi}) and the more detailed study of
Ref.~\cite{RCSI93}, which obtained \mbox{$f_\pi = 0.091$ GeV}, suggest that
meson-loop effects should not make a large contribution to the normalisation.
This is indeed what we find, as illustrated in Fig.~\ref{FigNorm} which shows
that for reasonable values of $M_c$ and $\Lambda_2$:
\begin{eqnarray}
\label{ParamV}
0.2\;{\rm GeV} < M_c < 0.6\;{\rm GeV}\;\; & \;\; {\rm and}\;\; &
\;\; 0.3\;{\rm GeV} < \Lambda_2 < 0.7\;{\rm GeV}
\end{eqnarray}
the correction is small and negative, leading to a less than 2\% correction
at the physical value of the pion mass.

The relative importance of the quark core and pion-loop contributions to the
electromagnetic pion charge radius, $r_\pi$, can also be obtained from
Eq.~(\ref{PiRpi}):
\begin{equation}
\label{rpiC}
\hat{r}_\pi^2 = 
-6 \frac{d}{d\,Q^2} \left.\ln \hat{F}_\pi(Q^2)\right|_{Q^2 = 0}~.
\end{equation}
Substituting we obtain
\begin{equation}
\hat{r}_\pi^2 = (r_\pi^2)^{\rm GIA} + \frac{16}{\pi^2\,\hat{f}_\pi^2}
        \,\int_0^\infty\,du\,\frac{u^3}{[4u + (m_\pi^{\rm L})^2]^4}
                                   \,g(u) 
\end{equation}
where $(r_\pi^2)^{\rm GIA}$ is the generalised impulse approximation
contribution. 

The result in Eq.~(\ref{fpirpi}) and that of \mbox{$r_\pi = 0.59$~fm}
obtained in Ref.~\cite{RCSI93} suggest that meson-loop effects can additively
contribute $10$-$15$\% to the charge radius at the physical value of the pion
mass.  The same conclusion can be drawn from the phenomenological Coulomb
gauge DSE studies of Ref.~\cite{LAA89}.  This is just what we find, as we
illustrate in Fig.~\ref{FigRad} which shows that with $\hat{f}_\pi=0.093$~GeV,
reasonable values of $M_c$ and $\Lambda_2$, Eq.~(\ref{ParamV}), and at the
physical value of the pion mass, the correction to the square of the charge
radius is always positive and less than 15\%.

It is of interest to study the chiral limit.  It will be recalled that we
have set the mass of all the external pions to zero so that the contribution
of the loop-pions in the chiral limit is easily identified:
\mbox{$m_\pi^{\rm L}\rightarrow 0$}.  As we have remarked, the generalised
impulse approximation contribution, $(r_\pi^{\rm GIA})$, is regular in the
chiral limit and only weakly dependent on the current quark mass.  In
Fig.~\ref{ChDiv} we plot the ratio \mbox{$\hat{r}_\pi^2/(r_\pi^{\rm GIA})^2$}
as a function of $m_\pi^{\rm L}$ for \mbox{$M_c = 0.22$ GeV} and
\mbox{$\Lambda_2 = 0.5$~GeV}.  This figure emphasises that at 
\mbox{$m_\pi^{\rm L} \approx 0.14$ GeV} the dominant contribution to the pion
charge radius is provided by the quark core.  It is not until $m_\pi^{\rm L}$
becomes very small, on the order of $0.01$~GeV, that the pion cloud
contribution becomes as important as the quark core and this contribution is
well described by the form
\begin{equation}
\label{lnfit}
(r_\pi^2)^{\rm div} = (r_\pi^2)^{\rm GIA} \left[
  0.73 - 0.082\,\ln\left(\frac{(m_\pi^{\rm L})^2}{m_\rho^2}\right)\right]~. 
\end{equation} 
for \mbox{$m_\pi^{\rm L} < 0.14$ GeV}. 

It is natural to compare our result with that of ChPT.  In Ref.~\cite{GL85}
one finds in Eq.~(5.6) the result
\begin{equation}
\langle r^2\rangle_\pi = \frac{12\,L_9^r}{F_0^2} 
        - \frac{1}{32\pi^2\,F_0^2} 
        \left( 2\,\ln\left[\frac{m_\pi^2}{\mu^2}\right] + 
                \ln\left[\frac{m_\kappa^2}{\mu^2}\right] + 3\right)
\end{equation}
where $\mu^2$ is the loop regularisation scale and $L_9^r$ is one of the ten
standard low energy constants in the effective action of ChPT into which an
infinity from the divergence of the pseudoscalar loop has been
absorbed\cite{fna}.  

The ten low energy coefficients, of which $L_9^r$ is one, are fixed by
requiring that the effective action of ChPT provide a good description of low
energy strong interaction phenomena while the loop regularisation scale
$\mu^2$ is usually taken to be something of the order of $m_\eta^2$ or
$m_\rho^2$ because this is a measure of the mass scale at which additional
excitations should be accounted for.  Of course, the actual values of the
coefficients $L_i^r$ depend on $\mu^2$ because of the ambiguities associated
with regularising the divergent loops but, once a scale is chosen, values can
be quoted.

Fitting $\kappa^0_{e3}$ decay using \mbox{$\mu^2 = m_\eta^2$}
[Ref.~\cite{GL85a}, below Eq.~(14.1)] and \mbox{$F_0 = 93.3$ MeV} a value of
\begin{equation}
L_9^r = (7.1 \,\pm \,0.4)\times\,10^{-3} 
\end{equation}
is quoted in Ref.~\cite{GL85} in which case
\begin{equation}
\frac{12\,L_9^r}{F_0^2} = (0.38\,\pm\,0.020)\;{\rm fm}^2~,
\end{equation}
which is 90\% of the experimental value: \mbox{$0.439\,\pm\,0.008$
fm$^2$}\cite{SRA86}.  A similar calculation in Ref.~\cite{UGM93} using
\mbox{$\mu^2 = m_\rho^2$} with
\begin{equation}
L_9^r = (6.9 \,\pm \,0.2)\times\,10^{-3} 
\end{equation}
yields [upon correction of a minor typographical error therein]
\begin{equation}
\frac{12\,L_9^r}{F_0^2} = (0.37\,\pm\,0.011)\;{\rm fm}^2~,
\end{equation}
which is 84\% of the experimental value.  One sees that at the physical value
of the pion mass, and with an accepted value of the renormalisation scale,
the ``chiral logarithm'' contributes little to the charge radius of the pion.

The pion charge radius has also been studied in nonlinear chiral theories,
Ref.~\cite{EV81}, where a comparison is made between the contribution of
nucleon and pion loops.  In this study the pion loop contribution at the
physical value of the pion mass was directly identified and found to be
\mbox{$0.06$ fm$^2$} which is just 14\% of the experimental value.

\section{Summary and Conclusions}
Herein we have calculated the charge radius of the pion in the coupled
Dyson-Schwinger--Bethe-Salpeter equation approach as formalised in the Global
Colour-symmetry model, which is a QCD based model field theory.  In this
approach, the ``bare hadrons'' are nonpointlike objects with a ``quark core''
obtained as the solution of Bethe-Salpeter (mesons) or relativistic Faddeev
(baryons) equations whose kernel is chosen in such a way as to model the
quark-quark interaction in QCD.  The meson cloud associated with a given
hadron is a small correction in this framework and is obtained through a
systematic expansion based on the effective action of the GCM.

In this approach the charge radius of the pion receives a contribution from
its quark core and from pion loops.  We showed that the quark core
contribution is finite in the chiral limit and that, at the physical value of
the pion mass, it is the dominant determining characteristic of the pion with
the pion loop contribution being a small, finite, additive correction of less
than 15\%.  The fact that the loop contribution is finite is a general
property of this formalism; it is due to the internal quark core structure
which provides a natural cutoff in all integrals that arise.  Our result is
consistent with a recent reanalysis of the pion charge radius obtained in
lattice simulations\cite{LC93}.

We explicitly identified the origin of the chiral divergence in the charge
radius of the pion as arising from the pion loop; this feature is not absent
in the general DSE-BSE approach but is simply an higher order correction, as
are all meson loop effects.  Our calculation demonstrated that the pion's own
pion cloud is not a very significant component unless the mass of the pion is
less than $\sim 50$ MeV.  

A broad implication of our result is that it demonstrates that the many
quark-based models which do not reproduce the logarithmic divergence of the
charge radius in the chiral limit; for example, the relativistic, constituent
quark model of Ref.~\cite{CCP88}, nevertheless capture the most important
characteristic of the pion: its quark core, and should not be discarded on
this basis alone.

In closing we note that the contribution of the quark core of the
proton/neutron to its charge radius has not been calculated in the approach
we have used herein, although such a calculation is underway.  In our
opinion, and if the analysis of Ref.~\cite{LC93} is a reasonable guide, the
pion loop contribution will be a little more important in this case but, at
the physical pion mass, will still generate less than a 30\% correction.

\acknowledgements
RA and AB take this opportunity to express their gratitude to Prof. Dr. H.
Reinhardt for his support and interest over the course of this work.  In
addition, AB would like to thank the Physics Division of Argonne National
Laboratory for their hospitality and for financial support during a visit in
which a significant part of this work was conducted; and also gratefully
acknowledges financial support from the Studienstiftung des deutschen Volkes.
CDR takes this opportunity to thank Prof. Dr. H. Reinhardt for his
hospitality during two visits to Universit\"{a}t T\"{u}bingen: one in which
this collaboration was formed and another in order to bring this work to a
conclusion. The authors also gratefully acknowledge useful discussions with
F. Coester. The work of CDR was supported by the US Department of Energy,
Nuclear Physics Division, under contract number W-31-109-ENG-38.  Some of the
calculations described herein were carried out using a grant of computer time
and the resources of the National Energy Research Supercomputer Center.

\appendix
\section*{Miscellaneous Expressions}
In this appendix we present expressions for $(r_{\pi}^2)^{\rm GIA}$,
Eq.~(\ref{AppRpi}), and $g(u)$, Eq.~(\ref{guExp}), since they are rather
lengthy.

The generalised impulse approximation contribution to the pion charge radius
is \cite{RCSI93}:
\begin{equation}
\label{AppRpi}
( r_{\pi}^2)^{\rm GIA}  = 
\frac{N_c}{2 f_\pi^2}\int\frac{d^4q}{(2\pi)^4}\, H(q^2)
\end{equation}
where
\begin{eqnarray}
H(x=q^2) & = & 
{\it B}\,\left (-{\it A}^{2}{\it B}^{2}{\it B'}\,{\it K''}\,{\it 
K}^{2}x^{2}-{\it A}^{4}{\it B''}\,{\it K'}\,{\it K}^{2}x^{3} -36\,
{\it A}^{2}{\it B}^{3}{\it K'}\,{\it K}^{2}\right. \nonumber \\
& & -{\it A}^{4}{\it B'}
\,{\it K''}\,{\it K}^{2}x^{3} +{\it A}^{2}{\it B}\,{\it A'}\,{\it 
A''}\,{\it K}^{3}x^{3}-30\,{\it A}\,{\it B}^{3}{\it A'}\,{\it K}
^{3} \nonumber \\
& & -42\,{\it A}^{2}{\it B}^{2}{\it B'}\,{\it K}^{3}-24\,{\it A}^
{2}{\it B}^{3}{\it K''}\,{\it K}^{2}x -48\,{\it A}^{4}{\it B}\,{
\it K'}\,{\it K}^{2}x\nonumber \\
& & -54\,{\it A}\,{\it B}^{3}{\it A'}\,{\it K'}
\,{\it K}^{2}x
-42\,{\it A}^{2}{\it B}^{2}{\it B'}\,{\it K'}\,{
\it K}^{2}x -72\,{\it A}^{3}{\it B}\,{\it A'}\,{\it K}^{3}x \nonumber \\
& & -18\,{
\it B}^{3}{\it A'}^{2}{\it K}^{3}x-24\,{\it A}^{4}{\it B'}\,{\it 
K}^{3}x-42\,{\it A}\,{\it B}^{2}{\it A'}\,{\it B'}\,{\it K}^{3}x
 \nonumber \\
& & -24\,{\it A}\,{\it B}^{3}{\it A''}\,{\it K}^{3}x -21\,{\it A}^{2}{
\it B}^{2}{\it B''}\,{\it K}^{3}x-4\,{\it A}^{2}{\it B}^{3}{\it 
K'}^{3}x^{2}\nonumber \\
& & +4\,{\it A}^{2}{\it B}^{3}{\it K''}\,{\it K'}\,{\it 
K}\,x^{2}-6\,{\it A}\,{\it B}^{3}{\it A'}\,{\it K'}^{2}{\it K}\,
x^{2} -2\,{\it A}^{2}{\it B}^{2}{\it B'}\,{\it K'}^{2}{\it K}\,x^{
2}\nonumber \\
& & -24\,{\it A}^{4}{\it B}\,{\it K''}\,{\it K}^{2}x^{2}-3\,{\it A}
\,{\it B}^{3}{\it A'}\,{\it K''}\,{\it K}^{2}x^{2}-68\,{\it A}^{3
}{\it B}\,{\it A'}\,{\it K'}\,{\it K}^{2}x^{2} \nonumber \\
& & -2\,{\it B}^{3}{
\it A'}^{2}{\it K'}\,{\it K}^{2}x^{2}-28\,{\it A}^{4}{\it B'}\,{
\it K'}\,{\it K}^{2}x^{2}-18\,{\it A}\,{\it B}^{2}{\it A'}\,{\it 
B'}\,{\it K'}\,{\it K}^{2}x^{2}\nonumber \\
& & -8\,{\it A}\,{\it B}^{3}{\it A''}
\,{\it K'}\,{\it K}^{2}x^{2} +3\,{\it A}^{2}{\it B}^{2}{\it B''}\,
{\it K'}\,{\it K}^{2}x^{2}-20\,{\it A}^{2}{\it B}\,{\it A'}^{2}{
\it K}^{3}x^{2}\nonumber \\
& & -40\,{\it A}^{3}{\it A'}\,{\it B'}\,{\it K}^{3}x^{
2}-14\,{\it A}\,{\it B}\,{\it A'}\,{\it B'}^{2}{\it K}^{3}x^{2}+2
\,{\it A}^{2}{\it B'}^{3}{\it K}^{3}x^{2}\nonumber \\
& & -36\,{\it A}^{3}{\it B}
\,{\it A''}\,{\it K}^{3}x^{2}-5\,{\it B}^{3}{\it A'}\,{\it A''}\,
{\it K}^{3}x^{2}-4\,{\it A}\,{\it B}^{2}{\it B'}\,{\it A''}\,{
\it K}^{3}x^{2} \nonumber \\
& & -8\,{\it A}^{4}{\it B''}\,{\it K}^{3}x^{2}+2\,{\it 
A}\,{\it B}^{2}{\it A'}\,{\it B''}\,{\it K}^{3}x^{2}+4\,{\it A}^
{2}{\it B}\,{\it B'}\,{\it B''}\,{\it K}^{3}x^{2}\nonumber \\
& & -4\,{\it A}^{4}{
\it B}\,{\it K'}^{3}x^{3}+4\,{\it A}^{4}{\it B}\,{\it K''}\,{\it 
K'}\,{\it K}\,x^{3}-6\,{\it A}^{3}{\it B}\,{\it A'}\,{\it K'}^{2
}{\it K}\,x^{3}\nonumber \\
& & -2\,{\it A}^{4}{\it B'}\,{\it K'}^{2}{\it K}\,x^{3
}-3\,{\it A}^{3}{\it B}\,{\it A'}\,{\it K''}\,{\it K}^{2}x^{3}-10
\,{\it A}^{2}{\it B}\,{\it A'}^{2}{\it K'}\,{\it K}^{2}x^{3}\nonumber \\
& & -10\,
{\it A}^{3}{\it A'}\,{\it B'}\,{\it K'}\,{\it K}^{2}x^{3}-4\,{
\it A}^{3}{\it B}\,{\it A''}\,{\it K'}\,{\it K}^{2}x^{3}-6\,{\it 
A}\,{\it B}\,{\it A'}^{3}{\it K}^{3}x^{3} \nonumber \\
&& \left. -6\,{\it A}^{2}{\it A'}
^{2}{\it B'}\,{\it K}^{3}x^{3}-2\,{\it A}^{3}{\it B'}\,{\it A''}
\,{\it K}^{3}x^{3}-2\,{\it A}^{3}{\it A'}\,{\it B''}\,{\it K}^{3}
x^{3} \right )~.
\label{ChRad}
\end{eqnarray}

As we remarked above, the integral in Eq.~(\ref{guint}) can be evaluated.
Writing $x=M_c^2/\Lambda_2^2$ and $\tilde{u}=u/\Lambda_2^2$ one obtains:
\begin{eqnarray}
\label{guExp}
g(u) & = & 
\frac{(x-1)^3}{((1+x)\ln x +2(1-x))(\tilde{u}-1+x)^3
[(\tilde{u}-1+x)^2+4\tilde{u}]} \times\\
     &   & \left\{ \rule{0mm}{8mm}(\tilde{u}+1-x)(\tilde{u}-1+x)^2
\right.\nonumber \\ 
     &  &
+\frac{4\tilde{u}x(\tilde{u}-1+x)+(\tilde{u}+1-x)(\tilde{u}+1+x)[(\tilde{u}-1+x)^2+4\tilde{u}]} 
                               {\sqrt{(\tilde{u}-1+x)^2+4\tilde{u}}}\nonumber\\ 
     &   & \times\ln
\left(\frac{\sqrt{(\tilde{u}-1+x)^2+4\tilde{u}}-(\tilde{u}+1-x)} 
                {(\tilde{u}+x)\sqrt{(\tilde{u}-1+x)^2+4\tilde{u}}+(\tilde{u}+x)^2+\tilde{u}-x}\right)\nonumber\\ 
     &  &
+\frac{[(\tilde{u}+1)(\tilde{u}-1+x)+(\tilde{u}+1-x)(\tilde{u}+1+x)][(\tilde{u}-1+x)^2+4\tilde{u}]} 
                  {2\sqrt{\tilde{u}(\tilde{u}+x)}}\nonumber\\  
     &   & \times\left. \ln
\left(\frac{\sqrt{\tilde{u}(\tilde{u}+x)}+\tilde{u}} 
                {\sqrt{\tilde{u}(\tilde{u}+x)}-\tilde{u}}\right)
\right\}.\nonumber 
\end{eqnarray}



\begin{figure}
\caption{This figure is a pictorial representation of the amplitude
identified with the $\pi\pi\gamma$ vertex in a generalised impulse
approximation.  The straight external lines represent the incoming and
outgoing $\pi$, the filled circles at the $\pi$ legs represent the
$\langle\pi|\overline{q}q\rangle$ Bethe-Salpeter amplitudes, the wiggly line
represents the photon, $\gamma$, the shaded circle at the $\gamma$ leg
represents the regular part of the dressed quark-photon vertex [which
satisfies the Ward-Takahashi Identity, Eq.~(\protect\ref{WTI})] and the
broken internal lines represent the dressed quark propagator.}
\label{figppA}
\end{figure}

\begin{figure}
\caption{This figure is a pictorial representation of our approximation to
the pion loop contribution to the $\pi\pi\gamma$ vertex,
Eq.~(\protect\ref{PionLoop}).  The straight external lines represent the
external pions, the large filled region at the $\pi$ legs represents
\mbox{$T^{ijkl}(p+q,p-k,k-q)$}, Eq.~(\protect\ref{Tijkl}),  the
internal lines represent the virtual loop pions, the smaller filled region
represents \mbox{$\Lambda_\mu^{\rm GIA}(k,p+q-k)$},
Fig.~\protect\ref{figppA}, and the wiggly line represents the photon. }
\label{figpiloop}
\end{figure}

\begin{figure}
\caption{In this figure we plot $[f_\pi - \hat{f}_\pi]$, in MeV, in order to
illustrate the dependence of the pion-loop induced renormalisation of $f_\pi$
on $M_c$ and $\Lambda_2$, physically reasonable ranges for which are
\mbox{$0.2\;{\rm GeV} < M_c < 0.6\;{\rm GeV}$} and
\mbox{$0.3\;{\rm GeV} < \Lambda_2 < 0.7\;{\rm GeV}$}.}
\label{FigNorm}
\end{figure}

\begin{figure}
\caption{In this figure we plot $[\hat{r}^2_\pi/(r^2_\pi)^{\rm GIA} - 1]$ in
order to illustrate the dependence of the pion loop contribution to $\langle
r^2_\pi\rangle$ on $M_c$ and $\Lambda_2$, physical ranges for which are
\mbox{$0.2\;{\rm GeV} < M_c < 0.6\;{\rm GeV}$} and
\mbox{$0.3\;{\rm GeV} < \Lambda_2 < 0.7\;{\rm GeV}$}.}
\label{FigRad}
\end{figure}

\begin{figure}
\caption{In this figure we plot $[\hat{r}^2_\pi/(r^2_\pi)^{\rm GIA} - 1]$ as
a function on $m_\pi$, in MeV, in order to illustrate the onset of the
pion-loop induced $\ln\,m_\pi$ divergence of $\langle \hat{r}^2_\pi\rangle$.
It is clear from this figure that even for $m_\pi = 50$~MeV the pion-loop
contributes $< 25$\% of the charge radius. The dashed line is the fit of
Eq.~(\protect\ref{lnfit}). }
\label{ChDiv}
\end{figure}


\begin{references}
\bibitem{EQnos} The relevant equations in Ref.~\cite{RCSI93} are (28),(29),
(33)-(37) and (39).
\bibitem{fna} One might make an operational identification of the $L_9^r$
term with the quark core contribution in our approach.  This might be argued
by interpreting the absorption of the infinity from the pion loop as a step
towards properly representing the nonpointlike nature of the physical
pseudoscalar mesons.
\end{references}
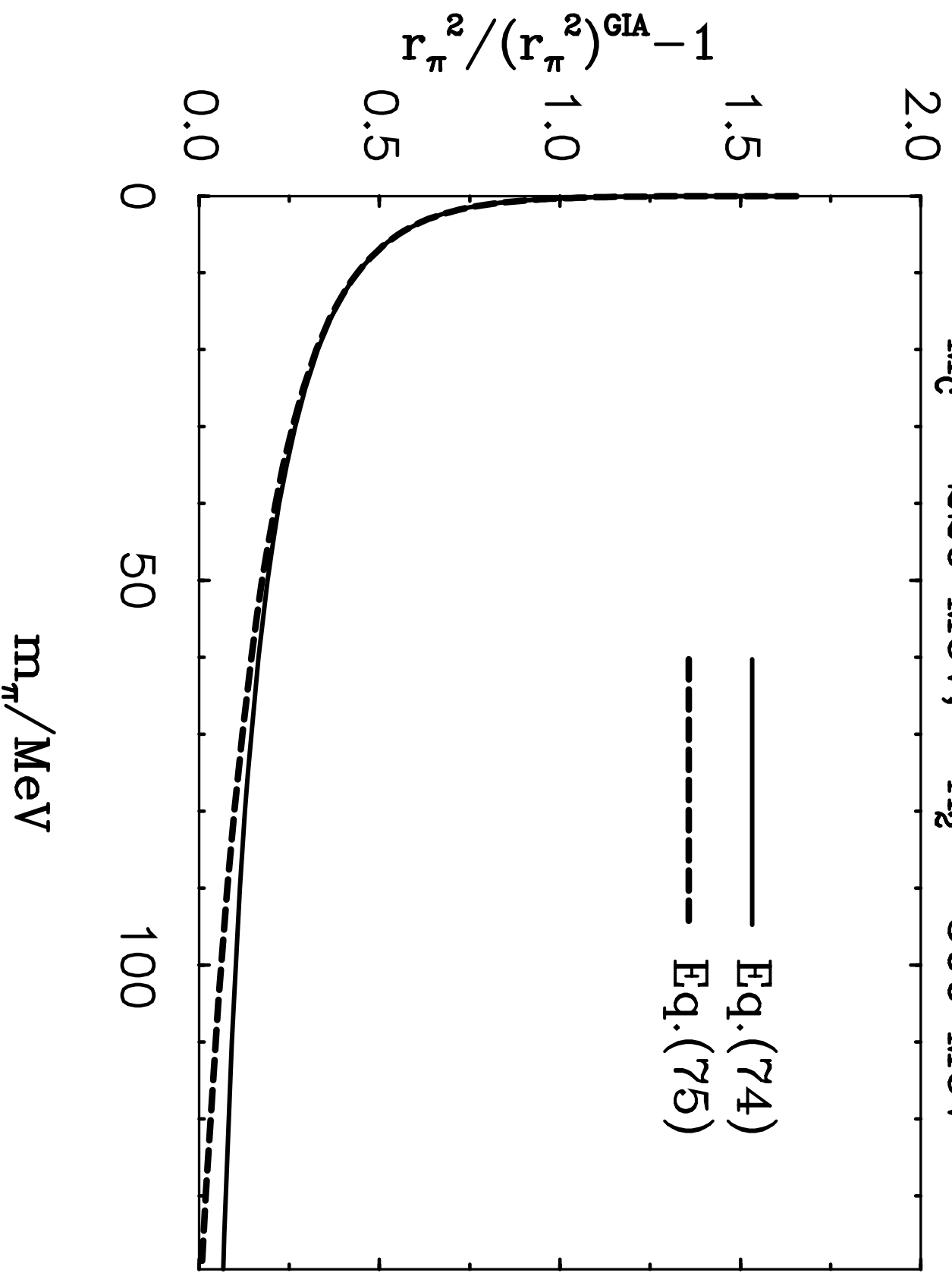
\end{document}